\documentclass[a4paper,twoside]{article}
\newcommand{\affil}[1]{$^{\rm #1}$}
\usepackage[authoryear]{natbib}
\bibpunct{(}{)}{;}{a}{}{,}
\usepackage{graphicx}
\date{} %Please leave the date blank
\newcommand{\Msun}{\ensuremath{\mathrm{M}_\odot}}
\newcommand{\Rsun}{\ensuremath{\mathrm{R}_\odot}}
\newcommand{\lsim}{\mathrel{\hbox{\rlap{\lower.55ex \hbox {$\sim$}}
 \kern-.3em \raise.4ex \hbox{$<$}}}}
\newcommand{\gsim}{\mathrel{\hbox{\rlap{\lower.55ex \hbox {$\sim$}}
 \kern-.3em \raise.4ex \hbox{$>$}}}}
\title{\large\bf\flushleft The Puzzling Frequencies of CEMP and NEMP Stars}
\author{\parbox{\textwidth}{\flushleft
\vspace{-0.5cm}
{\it O. R. Pols\affil{A}, R. G. Izzard\affil{A,B},
E. Glebbeek\affil{C} and R. J. Stancliffe\affil{D}
}\\
\vspace{0.4cm}
{\small \affil{A}\,Sterrekundig Instituut Utrecht, P.O. Box 80000, 
  NL-3584 TA Utrecht, The Netherlands. Email: O.R.Pols@uu.nl}\\
{\small \affil{B}\,Institut d'Astronomie et d'Astrophysique, 
  Universit\'e Libre de Bruxelles, CP226, Boulevard du Triomphe, 
  B-1050 Bruxelles, Belgium}\\
{\small \affil{C}\,Department of Physics and Astronomy, McMaster University,
  Hamilton, Ontario, L8S 4M1, Canada}\\
{\small \affil{D}\,Centre for Stellar and Planetary Astrophysics, Monash 
  University, P.O. Box 28M, Clayton, VIC 3800, Australia}
}}
\begin{document}
\twocolumn[
\begin{changemargin}{.8cm}{.5cm}
\begin{minipage}{.9\textwidth}
\vspace{-1cm}
\maketitle
%
%
%%%%%%%%%%%%%     ABSTRACT    %%%%%%%%%%%%%
%Abstract of no more than 200 words here.
\small{\bf Abstract:}
We present the results of binary population simulations of carbon- and
nitrogen-enhanced metal-poor (CEMP and NEMP) stars. We show that the
observed paucity of very nitrogen-rich stars puts strong constraints
on possible modifications of the initial mass function at low
metallicity.

%%%%%%%%%%%%%     KEYWORDS    %%%%%%%%%%%%%
\medskip{\bf Keywords:} 
stars: AGB and post-AGB, stars: evolution, stars: binaries, 
stars: abundances, stars: mass function
% Please write all keywords in lower case. PASA uses the
% standard list of subject headings adopted by The Astrophysical Journal
% and available from http://www.journals.uchicago.edu/ApJ/keywords_text.html.
% Keywords are separated by em-dashes, i.e. ---

%%%%%%%%DO NOT EDIT%%%%%%%%%%%%
\medskip
\medskip
\end{minipage}
\end{changemargin}
]
\small
%%%%%%%%EDIT FROM HERE%%%%%%%%%%%%

\section{Introduction}
%Please see the PASA Style Guide for help with correct layout for your manuscript.
%Examples of tables and figures are given below.

Carbon-enhanced metal-poor (CEMP) stars make up at least 10\,\% and
probably as much as 20--25\,\% of very metal-poor stars with
$\mathrm{[Fe/H]} < -2$ (\citealp{2006_Frebel,2006_Lucatello}). These
stars have $\mathrm{[C/Fe]} > 1.0$, and about 80\,\% of CEMP stars are
also enriched in $s$-process elements (CEMP-s stars;
\citealp{2007_Aoki}).  A likely formation scenario is pollution by
mass transfer from a more massive AGB companion in a binary system,
which has since become a white dwarf. Radial velocity monitoring
indeed suggests that probably all CEMP-s stars are binaries
\citep{2005_Lucatello_binfrac}.

Within the mass transfer scenario, the large proportion of CEMP-s
stars requires the existence of a sufficient number of binary systems
with primary components that have undergone AGB nucleosynthesis. In
recent studies (\citealp{2005_Lucatello_IMF,2007_Komiya}) it has been
argued that a different initial mass function (IMF) is therefore
needed at low metallicity, weighted towards intermediate-mass
stars. If true, this in turn has important consequences for the
chemical evolution of the halo and of other galaxies. However, the
model calculations on which these estimates are based still contain
many uncertainties regarding the evolution and nucleosynthesis of
low-metallicity AGB stars, the efficiency of mass transfer, and the
evolution of the surface abundances of the CEMP stars themselves
(e.g., see Izzard et al., this volume).

Apart from carbon, substantial enhancements of nitrogen with respect
to iron are common among CEMP stars, typically with $\mathrm{[C/N]} >
0$. Detailed AGB nucleosynthesis models of low initial mass ($<
2.5\,\Msun$) produce carbon, but do not produce nitrogen because it is
burned during helium shell flashes. On the other hand, AGB models of
higher mass convert the dredged-up carbon into nitrogen by CN cycling
at the bottom of the convective envelope (hot bottom burning,
HBB). The surface abundances of these more massive AGB stars approach
the CN equilibrium ratio of $\mathrm{[C/N]} \approx -2$. Detailed
evolution models of AGB stars \citep{2007_Karakas} indicate that,
while at solar metallicity HBB sets in at $\gsim 5\,\Msun$,
significantly lower masses are needed at low metallicity ($\gsim
2.7\,\Msun$ at $\mathrm{[Fe/H]}=-2.3$). One may thus expect a
population of so-called nitrogen-enhanced metal-poor (NEMP) stars,
with $\mathrm{[N/Fe]} > 0.5$ and $\mathrm{[C/N]} < -0.5$. Although a
few examples of such stars are known, mostly at
$\mathrm{[Fe/H]}<-2.9$, they appear to be very rare
\citep{2007_Johnson}.

In this contribution we show that the number of NEMP stars sets an
additional constraint on possible changes to the IMF at low
metallicity. In Sect.~2 we present results of our binary population
synthesis simulations, and in Sect.~3 we give our conclusions.

\section{Binary population nucleosynthesis modelling}

We have simulated populations of metal-poor halo stars in binary
systems using the rapid synthetic binary nucleosynthesis code of
\citet{2004_Izzard} and \citet{2006_Izzard}. The algorithm and the
assumptions made in these simulations are briefly summarized by Izzard
et al.\ (this volume). We adopt a metallicity $Z=10^{-4}$
($\mathrm{[Fe/H]} = -2.3$) and select stars with ages between 10 and
13.7\,Gyr (roughly corresponding to the age of the halo) and $\log g <
4.0$. Among this sample we designate as CEMP stars those with
$\mathrm{[C/Fe]} > 1.0$ and as NEMP stars those with $\mathrm{[N/Fe]}
> 0.5$ and $\mathrm{[C/N]} < -0.5$, following the definition of
\citet{2007_Johnson}. Note that these definitions partly overlap.
We adopt a binary fraction of 100\,\%; for different assumptions the
resulting numbers of CEMP and NEMP stars should simply be scaled with
the assumed binary fraction.

We compare our model results with the statistics of the SAGA database
of metal-poor stars \citep{2008_Suda}. We selected 375 stars from the
database in a metallicity range $\mathrm{[Fe/H]}=-2.3\pm0.5$ and $\log
g < 4.0$. Of these, 296 have a C abundance measurement and 69 classify
as CEMP stars, yielding a CEMP fraction of 18--23\,\%, while 88\,\% of
these classify as CEMP-s (defined as having $\mathrm{[Ba/Fe]} > 0.5$).
Only one star classifies as a NEMP star, giving a very small nominal
NEMP fraction of $\sim$0.3\,\% and a ratio of NEMP to CEMP stars of
$\sim$0.015. If we consider an extended metallicity range
$-4<\mathrm{[Fe/H]}<-2$ in order to improve the number statistics, we
find 7 NEMP stars and 78 CEMP stars, giving a NEMP fraction of about
1.5\,\% and a NEMP to CEMP ratio of 0.09, which we regard as an upper
limit to the NEMP/CEMP ratio at $\mathrm{[Fe/H]}\approx-2.3$. We note
that the ratio of NEMP to CEMP stars is a better constraint on our
model predictions, because it is independent of various model
uncertainties, most notably the binary fraction.

The SAGA database is by no means a statistically complete sample.
Bright giants are clearly overrepresented with respect to fainter
giants and turnoff stars, compared to what is expected from the
relative lifetimes of these phases. However, neither the observed
sample nor our model results show a strong dependence of the CEMP
fraction on evolution state (represented e.g.\ by the effective
gravity). Therefore the results we describe below are probably not
greatly affected by this selection effect.

\begin{table}[t]
\begin{center}
\caption[]{%
   Number fractions of CEMP and NEMP stars among halo stars at
   $\mathrm{[Fe/H]}=-2.3$.} \label{tab:results1}
\begin{tabular}{lccc}
\hline
model & CEMP & NEMP & NEMP/CEMP \\
\hline
1A & 2.30\,\% & 0.35\,\% & 0.15 \\
1B & 3.50\,\% & 0.71\,\% & 0.20 \\
1C & 3.15\,\% & 0.62\,\% & 0.20 \\
1D & 4.81\,\% & 1.35\,\% & 0.28 \\
1E & 13.5\,\% & 26.6\,\% & 2.0~ \\
\hline
2A & ~9.4\,\% & 0.34\,\% & 0.04 \\
2B & ~8.5\,\% & 0.70\,\% & 0.08 \\
2C & 10.5\,\% & 0.61\,\% & 0.06 \\
2D & 12.5\,\% & 1.33\,\% & 0.11 \\
2E & 15.8\,\% & 26.4\,\% & 1.7~ \\
%\hline
%3A & 15.0\,\% & ~0.47\,\% & 0.03 \\
%3B & 14.2\,\% & ~0.98\,\% & 0.07 \\
%3C & 17.1\,\% & ~0.85\,\% & 0.05 \\
%3D & 21.4\,\% & ~1.89\,\% & 0.09 \\
%3E & 47.1\,\% & 42.6~\,\% & 0.92 \\
\hline
\end{tabular}
%\medskip\\
%$^a$Table footnotes go here.\\
\end{center}
\end{table}

In Table~\ref{tab:results1} we give the number fractions of CEMP stars
and NEMP stars resulting from our models, for various assumptions
regarding the physical ingredients (models 1 and 2) and the initial
distributions of binary parameters (A--E). Model 1 is our default
model, where we parameterize third dredge-up (3DUP) according to
detailed models by \citet{2002_Karakas}, yielding C-rich AGB stars for
initial masses $>1.2$\,\Msun. In Model 2 we assume much more efficient
3DUP in low-mass AGB stars, applying the modifications described in
Izzard et al.\ (this volume). This results in almost all stars with
initial masses 0.85--2.7\,\Msun\ becoming C-rich AGB stars and thus
potential CEMP progenitors.

The adopted binary parameter distributions are as follows:
\begin{itemize}
\item[A.]{} 
Default model: the initial primary masses $M_1$ are distributed
according to the solar neighbourhood IMF as derived by
\citet{1993_Kroupa}, the initial separations are drawn from a flat
distribution in $\log a$ (with $a$ between 3 and $10^5\,\Rsun$) and
the initial mass ratios from a flat distribution in $q=M_2/M_1$.
\item[B.]{}
Primary masses are distributed as in Model A, but mass ratios and
orbital periods are drawn from the distributions derived by
\citet{1991_Duquennoy} for the local population of G dwarfs, i.e.\ a
log-normal period distribution with a broad peak at 170 years and a
mass-ratio distribution with a broad peak at $M_2/M_1=0.23$.
\item[C.]{}
Primary masses are taken from an alternative IMF for the solar
neighbourhood by \citet{1979_Miller}, described by a log-normal
distribution with a median mass of 0.1\,\Msun\ and a width $\Delta\log
M = 0.67$, while separations and mass ratios are distributed as in
Model A.
\item[D.]{}
As Model C, but with a modified log-normal IMF with a median mass of
0.79\,\Msun\ and $\Delta\log M = 0.5$, as proposed by
\citet{2005_Lucatello_IMF}.
\item[E.]{}
As Model C using a log-normal IMF with a much larger median
mass of 10\,\Msun\ and $\Delta\log M = 0.33$, as proposed by
\citet{2007_Komiya}.
\end{itemize}

The models with default binary parameter distributions (1A and 2A) are
among those presented by Izzard et al.\ (this volume). As discussed in
that paper, our default model (1A) fails to account for the large
observed CEMP frequency, while Model~2A leads to an increase of the
CEMP fraction to almost 10\,\%, much closer to but still short of the
observed value. In both models the small overall NEMP fraction is
consistent with the observations, but only Model~2A, with enhanced
dredge-up, can reproduce the observed small NEMP/CEMP ratio.

Distributions B and C give some insight into the dependence of our
results on uncertainties in the local (solar-neighbourhood) binary
parameter distributions. Model~1B leads to an increase by a factor of
1.5 in the number of CEMP stars and a factor of 2 in the number of
NEMP stars compared to Model~1A, as the peak in the
\citet{1991_Duquennoy} period distribution coincides with the period
range in which mass transfer is effective. On the other hand, in
Model~2B the number of CEMP stars is somewhat smaller than in Model~2A
because most CEMP stars now come from binaries with primary masses
0.85--1.2\,\Msun\ and $q>0.7$, which is disfavoured by the mass-ratio
distribution. Adoption of the \citet{1979_Miller} IMF (Model~1C and
2C) also gives somewhat higher CEMP and NEMP fractions than the
\citet{1993_Kroupa} IMF, but the effect is modest.

The results for distributions D and E show the effect of a modified
IMF. Model~1D assumes the IMF that \cite{2005_Lucatello_IMF} suggest
is required to reproduce the large CEMP fraction. The number of CEMP
stars increases by a factor 2, but still falls short of the observed
value. The discrepancy between our and Lucatello's results arises
mainly because in our default model the initial primary mass and
period range contributing to CEMP stars are smaller than they
assumed. With enhanced third dredge-up (Model~2D) the increase is only
a modest 30\,\%. Models~1D and 2D also show an increased NEMP
fraction, by a factor 4 over Model~A, which is the result of a larger
weight of intermediate-mass stars in this IMF. The NEMP/CEMP ratio is
too large to be compatible with observations.  This effect is much
more extreme when we assume the IMF suggested by \cite{2007_Komiya}
which has a median mass of $10\,M_\odot$.  Although this gives rise to
a substantial CEMP fraction, close to being compatible with
observations without the need to assume enhanced dredge-up (Model~1E),
the CEMP stars are outnumbered by NEMP stars by a factor of two. This
is clearly incompatible with the observed limits on the number
fraction of NEMP stars.

\section{Conclusions}

Our binary population synthesis models show that the paucity of NEMP
stars among metal-poor halo stars is incompatible with a strongly
modified IMF at low metallicity, heavily weighted towards
intermediate-mass stars. The observed limit on the ratio of NEMP to
CEMP stars constrains allowed changes in the IMF to a median mass of
at most $\sim$0.8\,\Msun, the effect of which on the frequency of CEMP
stars is modest. A more promising explanation for both the ubiquity of
CEMP-s stars and the near-absence of NEMP stars is that low-mass,
low-metallicity AGB stars undergo much more efficient dredge-up of
carbon than shown by detailed evolution models available to date (also
see Izzard et al., this volume), perhaps as a result of proton
ingestion during the first thermal pulse (Cristallo et al., this
volume).

\section*{Acknowledgments}

We thank Falk Herwig, Maria Lugaro, Amanda Karakas and Selma de Mink
for enlightening discussions and useful feedback on the results
presented here. RGI's work in Utrecht is supported by the NWO under
grant 614.000.424. RJS is funded by the Australian Research Council's
Discovery Projects scheme under grant \\ DP0879472.

%\end{multicols}

\end{document}